\documentclass[twocolumn]{aastex631}
\usepackage[fleqn]{amsmath}
\usepackage{mathrsfs}
\usepackage{soul}
\usepackage{xcolor}
\usepackage{ctable}
\usepackage{graphicx}
\usepackage{dcolumn}
\usepackage{bm}
\usepackage{empheq}
\shorttitle{Moonraker}
\shortauthors{Mousis et al.}
\graphicspath{{./}{figures/}}

\begin{document}

\title{Moonraker -- Enceladus Multiple Flyby Mission}

\correspondingauthor{Olivier Mousis}
\email{olivier.mousis@lam.fr}

\author[0000-0001-5323-6453]{O. Mousis}
\affiliation{Aix-Marseille Universit\'e, CNRS, CNES, Institut Origines, LAM, Marseille, France}
\affiliation{Institut universitaire de France (IUF)}

\author{A. Bouquet}
\affiliation{Aix-Marseille Université, CNRS, Institut Origines, PIIM, Marseille, France}
\affiliation{Aix-Marseille Universit\'e, CNRS, CNES, Institut Origines, LAM, Marseille, France}

\author{Y. Langevin}
\affiliation{IAS, Institut d'Astrophysique Spatiale, Université Paris-Saclay, CNRS, Orsay, France}

\author[0000-0001-8017-5676]{N. Andr\'e}
\affiliation{Institut de Recherche en Astrophysique et Plan\'etologie, 9 avenue du Colonel Roche, 31028, Toulouse Cedex 4, France}

\author{H. Boithias}
\affiliation{Airbus Defence \& Space}

\author{G. Durry}
\affiliation{Groupe de Spectrométrie Mol\'eculaire et Atmosph\'erique, UMR 7331, CNRS, Universit\'e de Reims, \\
Champagne Ardenne, Campus Sciences Exactes et Naturelles, BP 1039, 51687, Reims, France}

\author{F. Faye}
\affiliation{Airbus Defence \& Space}

\author{P. Hartogh}
\affiliation{Max-Planck-Institut für Sonnensystemforschung, G\"ottingen, Germany}

\author[0000-0001-5346-9505]{J. Helbert}
\affiliation{Institute for Planetary Research, DLR, Berlin, Germany}

\author{L. Iess}
\affiliation{Dipartimento di ingegneria meccanica e aerospaziale, Universit\'a La Sapienza, Roma, Italy}

\author{S. Kempf}
\affiliation{LASP, University of Colorado, Boulder, CO, USA}

\author{A. Masters}
\affiliation{Imperial College London, London, UK}

\author{F. Postberg}
\affiliation{Institute of Geological Sciences, Freie Universit\"at Berlin, Germany}

\author{J.-B. Renard}
\affiliation{LPC2E, CNRS, Université Orl\'eans, CNES, 3A avenue de la Recherche Scientifique, 45071, Orl\'eans Cedex 2, France}

\author{P. Vernazza}
\affiliation{Aix-Marseille Universit\'e, CNRS, CNES, Institut Origines, LAM, Marseille, France}

\author[0000-0002-7400-9142]{A. Vorburger}
\affiliation{Physics Institute, University of Bern, Bern, Switzerland}

\author[0000-0002-2603-1169]{P. Wurz}
\affiliation{Physics Institute, University of Bern, Bern, Switzerland}

\author{D.H. Atkinson}
\affiliation{Jet Propulsion Laboratory, California Institute of Technology, 4800 Oak Grove Dr, Pasadena, CA 91109-8001, USA}

\author{S. Barabash}
\affiliation{Swedish Institute of Space Physics, Kiruna, Sweden}

\author{M. Berthomier}
\affiliation{Laboratoire de Physique des Plasmas, Ecole Polytechnique, Palaiseau, France}

\author{J. Brucato}
\affiliation{INAF-Astrophysical Observatory of Arcetri, Florence, Italy}

\author[0000-0002-3680-302X]{M. Cable}
\affiliation{Jet Propulsion Laboratory, California Institute of Technology, 4800 Oak Grove Dr, Pasadena, CA 91109-8001, USA}

\author{J. Carter}
\affiliation{Aix-Marseille Universit\'e, CNRS, CNES, Institut Origines, LAM, Marseille, France}

\author{S. Cazaux}
\affiliation{Faculty of Aerospace Engineering, Delft University of Technology, Delft, The Netherlands}
\affiliation{Leiden Observatory, Leiden University, PO Box 9513, 2300, RA Leiden, The Netherlands}

\author[0000-0003-3414-3491]{A. Coustenis}
\affiliation{LESIA, Paris Observatory, CNRS, PSL Universit\'e, Sorbonne Universit\'e, Universit\'e de Paris, 92190 Meudon, France}

\author{G. Danger}
\affiliation{Aix-Marseille Université, CNRS, Institut Origines, PIIM, Marseille, France}
\affiliation{Institut universitaire de France (IUF)}

\author[0000-0002-9516-8572]{V. Dehant}
\affiliation{Royal Observatory of Belgium, Brussels, 3 Avenue Circulaire, B1180, Brussels, Belgium}

\author{T. Fornaro}
\affiliation{INAF-Astrophysical Observatory of Arcetri, Florence, Italy}

\author{P. Garnier}
\affiliation{Institut de Recherche en Astrophysique et Planétologie, 9 avenue du Colonel Roche, 31028, Toulouse Cedex 4, France}

\author[0000-0002-9794-5056]{T. Gautier}
\affiliation{LATMOS-IPSL, CNRS, Sorbonne Université, UVSQ-UPSaclay, Guyancourt, France}
\affiliation{LESIA, Paris Observatory, CNRS, PSL Universit\'e, Sorbonne Universit\'e, Universit\'e de Paris, 92190 Meudon, France}

\author{O. Groussin}
\affiliation{Aix-Marseille Universit\'e, CNRS, CNES, Institut Origines, LAM, Marseille, France}

\author[0000-0002-8587-0202]{L.Z. Hadid}
\affiliation{Laboratoire de Physique des Plasmas (LPP), CNRS, Observatoire de Paris, Sorbonne Universit\'e, Universit\'e Paris Saclay, Ecole polytechnique, Institut Polytechnique de Paris, 91120 Palaiseau, France}

\author{J.-C. Ize}
\affiliation{Aix-Marseille Universit\'e, CNRS, CNES, Institut Origines, LAM, Marseille, France}

\author[0000-0002-1704-3846]{I. Kolmasova}
\affiliation{Department of Space Physics, Institute of Atmospheric Physics of the Czech Academy of Sciences, 141 00 Prague, Czech Republic}
\affiliation {Faculty of Mathematics and Physics, Charles University, Prague, Czech Republic}

\author{J.-P. Lebreton}
\affiliation{LPC2E, CNRS, Université Orl\'eans, CNES, 3A avenue de la Recherche Scientifique, 45071, Orl\'eans Cedex 2, France}

\author[0000-0002-9524-9479]{S. Le Maistre}
\affiliation{Royal Observatory of Belgium, Brussels, 3 Avenue Circulaire, B1180, Brussels, Belgium}

\author{E. Lellouch}
\affiliation{LESIA, Paris Observatory, CNRS, PSL Universit\'e, Sorbonne Universit\'e, Universit\'e de Paris, 92190 Meudon, France}

\author{J.I. Lunine}
\affiliation{Department of Astronomy, Cornell University, Ithaca, NY, USA}

\author{K.E. Mandt}
\affiliation{Johns Hopkins Applied Physics Laboratory, Laural, MD, USA}

\author[0000-0002-5420-1081]{Z. Martins}
\affiliation{Centro de Química Estrutural, Institute of Molecular Sciences and Department of Chemical Engineering,\\ 
Instituto Superior T\'ecnico, Universidade de Lisboa, Av. Rovisco Pais 1, 1049-001 Lisboa, Portugal}

\author[0000-0002-3427-2974]{D. Mimoun}
\affiliation{ISAE-SUPAERO, DEOS/SSPA, Universit\'e de Toulouse, France}

\author{Q. Nenon}
\affiliation{Institut de Recherche en Astrophysique et Planétologie, 9 avenue du Colonel Roche, 31028, Toulouse Cedex 4, France}

\author{G.M. Mu\~noz Caro}
\affiliation{Centro de Astrobiolog\'ia (INTA-CSIC), Ctra. de Ajalvir, km 4, E-28850 Torrej\'on de Ardoz, Madrid, Spain}

\author{P. Rannou}
\affiliation{Groupe de Spectrométrie Mol\'eculaire et Atmosph\'erique, UMR 7331, CNRS, Universit\'e de Reims, \\
Champagne Ardenne, Campus Sciences Exactes et Naturelles, BP 1039, 51687, Reims, France}

\author{H. Rauer}
\affiliation{Institute for Planetary Research, DLR, Berlin, Germany}

\author{P. Schmitt-Kopplin}
\affiliation{Max Planck Institute for Extraterrestrial Physics, MPI/MPE, Center for Astrochemical Studies, Garching, Germany}

\author{A. Schneeberger}
\affiliation{Aix-Marseille Universit\'e, CNRS, CNES, Institut Origines, LAM, Marseille, France}

\author{M. Simons}
\affiliation{Jet Propulsion Laboratory, California Institute of Technology, 4800 Oak Grove Dr, Pasadena, CA 91109-8001, USA}

\author{K. Stephan}
\affiliation{Institute for Planetary Research, DLR, Berlin, Germany}

\author{T. Van Hoolst}
\affiliation{Royal Observatory of Belgium, Brussels, 3 Avenue Circulaire, B1180, Brussels, Belgium}

\author{J. Vaverka}
\affiliation{Faculty of Mathematics and Physics, Charles University, Prague, Czech Republic}

\author[0000-0002-1760-210X]{M. Wieser}
\affiliation{Swedish Institute of Space Physics, Kiruna, Sweden}

\author{L. W\"orner}
\affiliation{German Aerospace Center, Institute for Quantum Technologies, Ulm, Germany}



\begin{abstract}

Enceladus, an icy moon of Saturn, possesses an internal water ocean and jets expelling ocean material into space. Cassini investigations indicated that the subsurface ocean could be a habitable environment having a complex interaction with the rocky core. Further investigation of the composition of the plume formed by the jets is necessary to fully understand the ocean, its potential habitability, and what it tells us about Enceladus’ origin. {\it Moonraker} has been proposed as an ESA M-class mission designed to orbit Saturn and perform multiple flybys of Enceladus, focusing on traversals of the plume. The proposed {\it Moonraker} mission consists of an ESA-provided platform, with strong heritage from JUICE and Mars Sample Return, and carrying a suite of instruments dedicated to plume and surface analysis. The nominal {\it Moonraker} mission has a duration of $\sim$13.5 years. It includes a 23--flyby segment with 189 days allocated for the science phase, and can be expanded with additional segments if resources allow. The mission concept consists in investigating: i) the habitability conditions of present-day Enceladus and its internal ocean, ii) the mechanisms at play for the communication between the internal ocean and the surface of the South Polar Terrain, and iii) the formation conditions of the moon. {\it Moonraker}, thanks to state-of-the-art instruments representing a significant improvement over Cassini’s payload, would quantify the abundance of key species in the plume, isotopic ratios, and physical parameters of the plume and the surface. Such a mission would pave the way for a possible future {landed mission}.

\end{abstract}

\keywords{Saturnian satellites --- Enceladus --- Astrobiology --- Natural satellite formation --- Natural satellite surfaces --- Natural satellite dynamics}


\section{Introduction} 
\label{sec:intro}

One of the most striking discoveries of the Cassini mission to the Saturn system is the direct observation of a plume \citep{Do06,To06,Wa06,Sp06,Ha06,Po06} emanating from Enceladus, a small (252 km radius) icy moon of Saturn, with multiple lines of evidence suggesting the plume is sourced from a subsurface liquid water ocean. The plume emanates from the geologically young South Polar Terrain (SPT) \citep{Po06,Do06,hansen2006enceladus,spahn2006cassini,tokar2006interaction,waite2006cassini} through four main fissures dubbed the ``Tiger Stripes''. It consists mostly of water vapor and water ice grains, as well as CO$_2$, CH$_4$, H$_2$, NH$_3$ and complex organics \citep{waite2006cassini,Hs15,Wa17,Po18,Gl20}. The $\sim$100 jets from the surface \citep{Po14}, and a likely more diffuse emission \citep{Sp15}, are responsible for the replenishment of Saturn’s diffuse E-ring and the Enceladus water torus \citep{Ha11}. 

\begin{figure}[!ht]
\resizebox{\hsize}{!}{\includegraphics[angle=0,width=5cm]{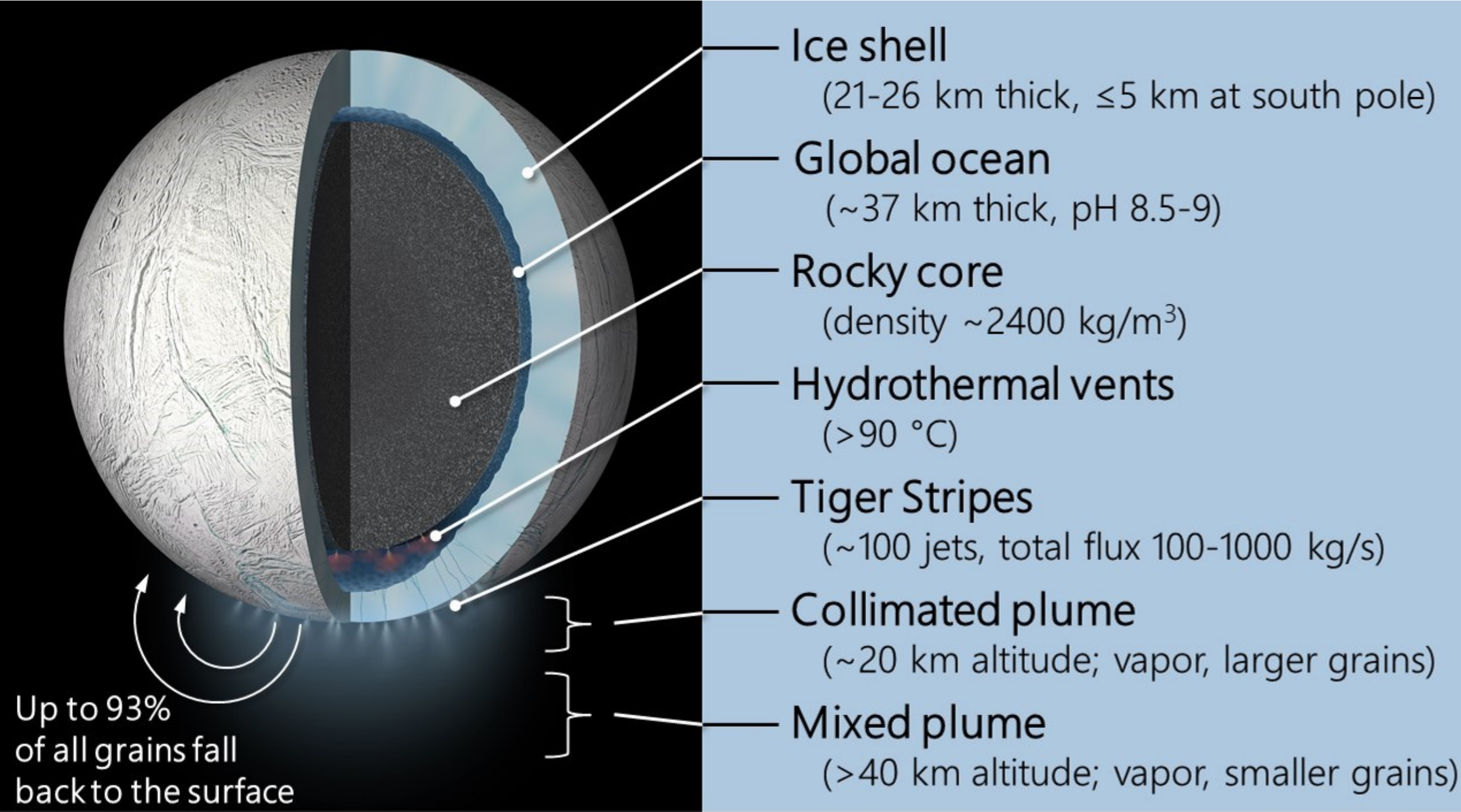}}
\caption{Main characteristics of Enceladus and its plume as understood currently. Reproduced from \cite{Ca21}, under CC BY license (https://creativecommons.org/licenses/by/4.0/). Background image: PIA20013 (NASA/JPL-Caltech). }
\label{fig:fig1}
\end{figure}

Cassini data revealed Enceladus as one of the most promising objects for habitability in the solar system \citep{He19,Ha22,Ca21}. The composition of the icy grains, containing various salts \citep{Po09,Po11}, demonstrates a subsurface liquid source for the plume material i.e., an internal sea that is in contact with a rocky core. Gravity measurements \citep{He19} and observations of the libration \citep{Th16} of the ice shell point to a global ocean under an icy crust of variable thickness (thinner under the SPT, see Fig. \ref{fig:fig1}), in contact with a rocky core of modest density, made of porous rock and/or aqueously altered minerals. Detection of H$_2$ in the gas phase of the plume \citep{Wa17}, and of SiO$_2$ particles in E-ring grains \citep{Hs15}, are evidence of ongoing or geologically recent hydrothermal activity on Enceladus’ seafloor, resulting from alteration of minerals by water. 

On Earth, seafloor hydrothermal vents provide chemical gradients that sustain lifeforms in the absence of sunlight. On Enceladus, the composition of the volatile phase of the plume indicates that its subsurface ocean features chemical disequilibria that would be usable for metabolic reactions \citep{Wa17,Ra21,Ho22}. Mass spectrometry measurements in the plume also show the presence of various organic molecules over a wide range of masses (e.g., tentative detection of C$_2$H$_6$, CH$_3$OH, C$_8$H$_{18}$; varied macromolecular organics with mass $>$ 200 amu, awaiting further characterization) \citep{Ma17,Po18}. Of the six elements commonly thought to be necessary for life (C, H, O, N, P, S), only P and S have not been firmly detected at Enceladus, likely due to their modest abundance and the limitations of Cassini’s instruments. The discovery of a strong thermal anomaly on Enceladus icy surface, and the evidence for hydrothermal chemistry (hydrogen gas, {silica nanoparticles}, sodium salt rich ice grains in the plume) along with the organics and liquid water, suggest that habitable conditions could exist beneath the moon's icy crust.

The question of the age and formation scenario of Enceladus is also not completely solved \citep{mckinnon2018mysterious}, with lines of evidence indicating it may be as recent as 200 Myr \citep{Cu16,Ne19}. This has implications for the formation timeline of the Saturnian system, the materials available in the ocean, the extent of the rock-water interaction, and the lifetime of the ocean – and therefore the time available for life to emerge.

\begin{figure}[!ht]
\resizebox{\hsize}{!}{\includegraphics[angle=0,width=5cm]{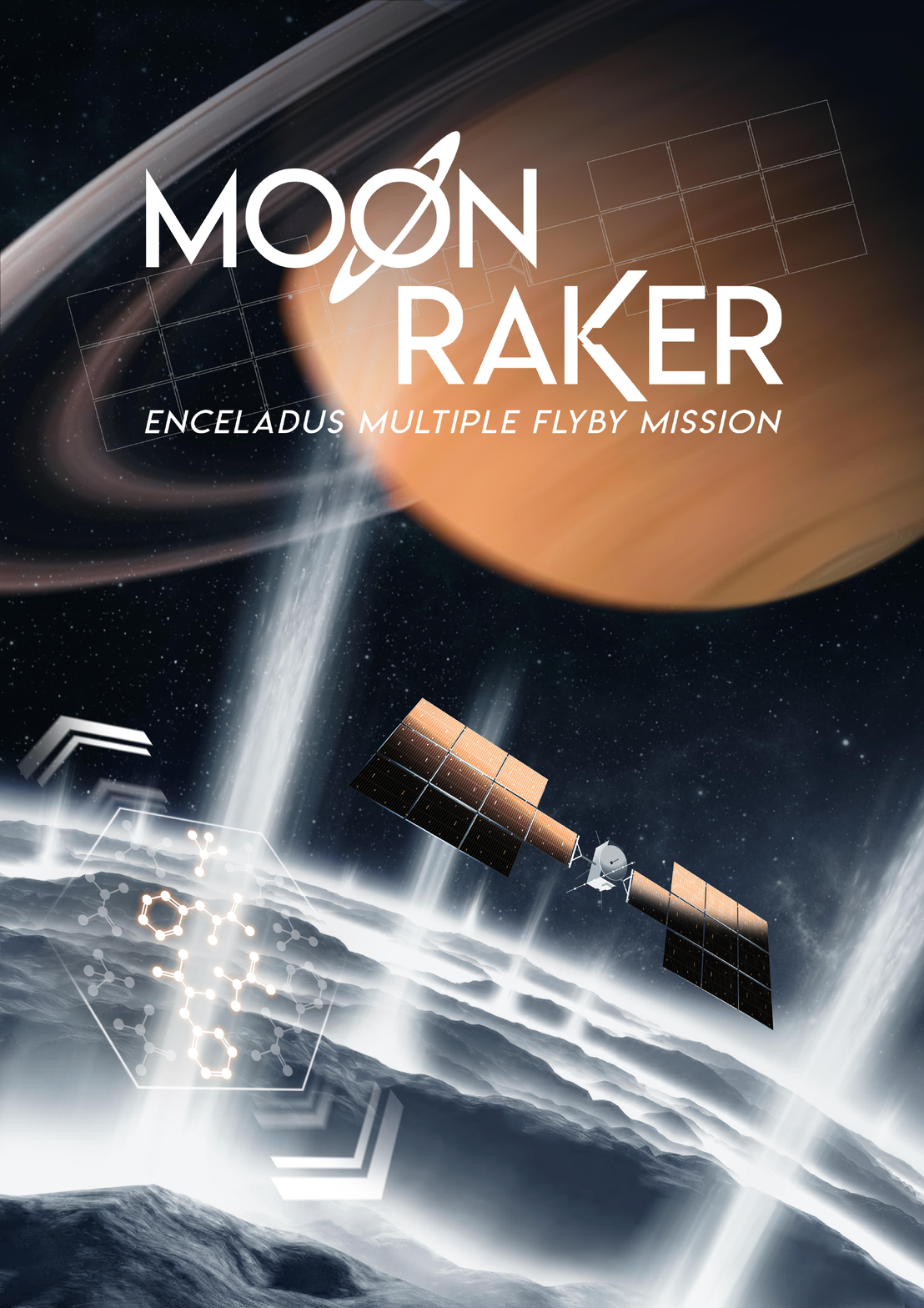}}
\caption{Artist view of the {\it Moonraker} mission concept. {Note that the geometry adopted for the image does not reflect reality. Enceladus orbits in the ring plane. The plume sources are at most 20$^\circ$ of the south pole \citep{Po14} and do not extend as far as 45$^\circ$ as suggested by the rendition.}}
\label{fig:fig2}
\end{figure}

All these discoveries and the related outstanding questions have recently led the {2023--2032 US National Academies' Planetary Science and Astrobiology Decadal Survey to recommend} NASA either {a large (Flagship) or midsize (New Frontiers) mission} whose aim would be to accomplish multiple flybys of Enceladus\footnote{https://www.nationalacademies.org/our-work/planetary-science-and-astrobiology-decadal-survey-2023-2032}, {prior to landing} on its surface in the case of the Flagship mission, in addition to a higher priority Flagship mission toward the Uranus system. In this paper, we describe a proposal for a Saturn orbiter aiming at accomplishing several dozens of flybys of Enceladus' SPT, and called the {\it Moonraker} mission (see Fig. \ref{fig:fig2}). This proposal has been submitted in response to the European Space Agency (ESA) Call for a medium-size mission opportunity (M--class Call) released at the end of 2021. It has been constrained by several limitations imposed by the M--class Call (limited budget, use of Ariane 62 (AR62), 12--year cruise duration, limited international collaboration, and limited operation duration). The submission of a more ambitious version of this proposal is now envisaged to the upcoming ESA Call for a large-size mission opportunity (L--class Call). 

The paper is organized as follows. Section \ref{sec:goals} summarizes the science goals of the {\it Moonraker} mission concept. The proposed mission and payload configurations are presented in Sec. \ref{sec:sec3} and Sec. \ref{sec:sec4}, respectively. The management structure of the proposed mission is detailed in Sec. \ref{sec:sec5}. Section \ref{sec:sec6} is devoted to conclusions and prospects.

\section{Science Goals}
 \label{sec:goals}
 
Through multiple flybys above the South Polar Terrain and both {\it in situ} and remote measurements, the {\it Moonraker} mission concept addresses three main science goals. Table \ref{tab:tab1} presents the Science Traceability Matrix.

\subsection{Science Goal 1 – Habitability conditions of present-day Enceladus}
Liquid water is only one of the conditions for life as we know it: metabolic energy and proper ``building blocks'' also need to be available in sufficient concentrations. The interface of the rocky core with the ocean is the most likely environment to be habitable on Enceladus; the composition of the ocean and therefore of the plume reflects the rock-water interaction. The measurements to be performed by {\it Moonraker} answer the following questions:

\begin{itemize}
\item \textit{How much chemical energy is available in the subsurface?} 
Without significant solar energy, complex chemical systems or even a hypothetical biosphere in Enceladus’ ocean would have to rely on chemical energy. Cassini’s measurements already indicated that methanogenesis is a viable reaction to sustain a potential biosphere \citep{Wa17,Hs15}, but other yet undetected chemical species could be used for other metabolic reactions \citep{Ra21}. To quantify this energy, the {\it Moonraker} mission would be able to measure the abundances in the plume (best proxy for the ocean itself) of key chemical species that could be used in metabolic reactions (e.g., CO$_2$, H$_2$, H$_2$S, sulfates, O$_2$). The flux of charged energetic particles from Saturn's magnetosphere hitting Enceladus' surface also contributes to form oxidants \citep{Te17}, that can be delivered to the reducing ocean due to constant burial by plume deposition that exceeds radiolytic destruction \citep{southworth2019surface,Ra21}; this delivery would help produce redox gradients.  {\it Moonraker} would perform {\it in situ} measurements of the {charged particle environment} in the vicinity of Enceladus, allowing for calculation of the amount of oxidants thus formed on the surface.   

\item \textit{Are the elements necessary for life as we know it (CHNOPS) present, and in what forms and abundances?}
While C, H, N, and O have been identified in Enceladus’ plume, P-bearing and S-bearing species are only tentatively detected, as phosphine and hydrogen sulfide respectively, in the gas phase of the plume \citep{Ma17}. Phosphorus and sulfur may also be present in other chemical forms, such as phosphates \citep{Ha22} and sulfates, which would be refractory. {Next} generation instruments aboard {\it Moonraker} would be capable of detecting trace abundances of these key species in the plume, whether in vapor or solid phase. 

\item \textit{What is the nature and extent of the interaction between the ocean and the rocky core?}
 Interaction between the ocean and the core not only provides chemical energy \citep{Wa17,Hs15,Bo17}; it can also produce a variety of organic compounds, ionic species, and heat to sustain the ocean over the age of Enceladus \citep{Ch17}. It is not clear at which stage of aqueous alteration the core currently is \citep{zandanel2021short}. Ocean-core interaction is also a key part in regulating the pH of the ocean \citep{glein2018geochemistry,Gl20}. Key measurements to understand the geochemistry of Enceladus include Ca, Mg, sulfates, and silica in icy grains. Organic compounds, in both gas and solid phase, represent also a critical target since they may reflect both Enceladus' initial inventory (leached by the ocean) or compounds synthesized through hydrothermal activity. Measurements in the plume can remove ambiguities left by the Cassini measurements and allow us to detect low abundance compounds of interest, including putative biomolecules.
\end{itemize}

\subsection{Science Goal 2 -- Communication between the subsurface ocean and the surface through the south polar terrain}
 There are still many standing questions about the exact mechanisms generating the plume, and their stability over geological timescales \citep{spencer2018plume}. The exact form of the emission (distribution of vapor and icy grains between the jets \citep{Po14} and the more diffuse, “curtain” source \citep{Sp15}), is still not fully determined. New observational constraints are required to understand what regulates the plume, possibly including opening and closing of vents \citep{In16,In17}. Aside from the intrinsic interest in understanding the generation of the plume, it is a crucial piece of context to interpret the data gathered from its composition and establish a link with oceanic composition. A deeper understanding of the SPT and its evolution with time is invaluable for possible future {landed missions} (e.g., \cite{Ma21}). The {\it Moonraker} mission concept addresses the following questions related to the plume generation:

\begin{itemize}
\item \textit{What is the size, distribution, and shape of vents and fractures from which the plume emanates, and the temperature distribution?} 
Our understanding of the morphology and size of the vents is limited by the capabilities of the Cassini spacecraft payload. Observations have provided only upper limits for vents size. Different models predict different channel widths, with different predictions such as vents shutting off after a few years \citep{In16} or a near insignificant endogenic heat emission between the fractures \citep{kite2016sustained}.  The study of the morphology of the vents and the associated heat flow (including the likely background emission between the stripes) would allow us to constrain plume emission models. The variation in width of the stripes (spatial variation as well as opening/closing with time) needs to be observed at high resolution; models predict the evolution of these openings on the timescale of years \citep{spencer2018plume}. The total heat flow and its distribution would allow us to quantify the effect of condensation in the vents. {\it Moonraker} would perform remote observations at a high spatial resolution to draw a new picture of the emission zone.

\item \textit{How are vapor and ice grains distributed in the plume? How has the plume evolved since the Cassini measurements?} 
The distribution and velocity of the gas phase and ice grains, as well as the size distribution of the ice grains, are critical parameters that models of plume generation must reproduce. The spatial distribution of vapor and solid sources (and correspondence to surface features) is another outstanding question \citep{spencer2018plume}. {\it Moonraker} would combine {\it in situ} and remote observations across many flybys to understand the plume production mechanism {and its evolution with time}. 
\end{itemize}

\subsection{Science Goal 3 -- Origin of Enceladus in the context of the formation of Saturn’s system} 
Possible formation scenarios for Enceladus include a primordial formation along with the other satellites, late formation from ring materials, or recent (100--200 Myr ago) formation due to a catastrophic event in the Saturnian  system \citep{mckinnon2018mysterious}. Distinction between these scenarios requires quantification of various tracers such as noble gases and their isotopic ratios, and their comparison to known measurements in the other satellites and objects indicative of the early solar system (e.g., chondrites and comets). Their quantification with {\it Moonraker} measurements would allow us to address the following question related to Enceladus’ origin: 

\begin{itemize}
\item \textit{Which volatile tracers are primordial, and which are evolved? How do the tracers compare to early solar system objects and other Saturnian system bodies? How old is Enceladus?} 
Noble gases abundances and isotopic ratios of noble gases, carbon, nitrogen, oxygen, and hydrogen are all tied to the reservoir from which the building blocks of Enceladus came. Comparison of these data with those already measured by Cassini-Huygens in Titan (Ar, Kr, and Xe abundances, D/H in CH$_4$, and $^{14}$N/$^{15}$N) would indicate if the building blocks of the two moons originate from the same material reservoir or if they followed distinct formation paths. {\it Moonraker} would perform measurements at Enceladus to remove the existing ambiguities and detect species that were under the limits of detection (LODs) of Cassini.
\end{itemize}

\begin{table*}
\caption{Moonraker Science Traceability Matrix}
\begin{tabular}{c}
\label{tab:tab1}
\includegraphics[angle=0,width=16cm]{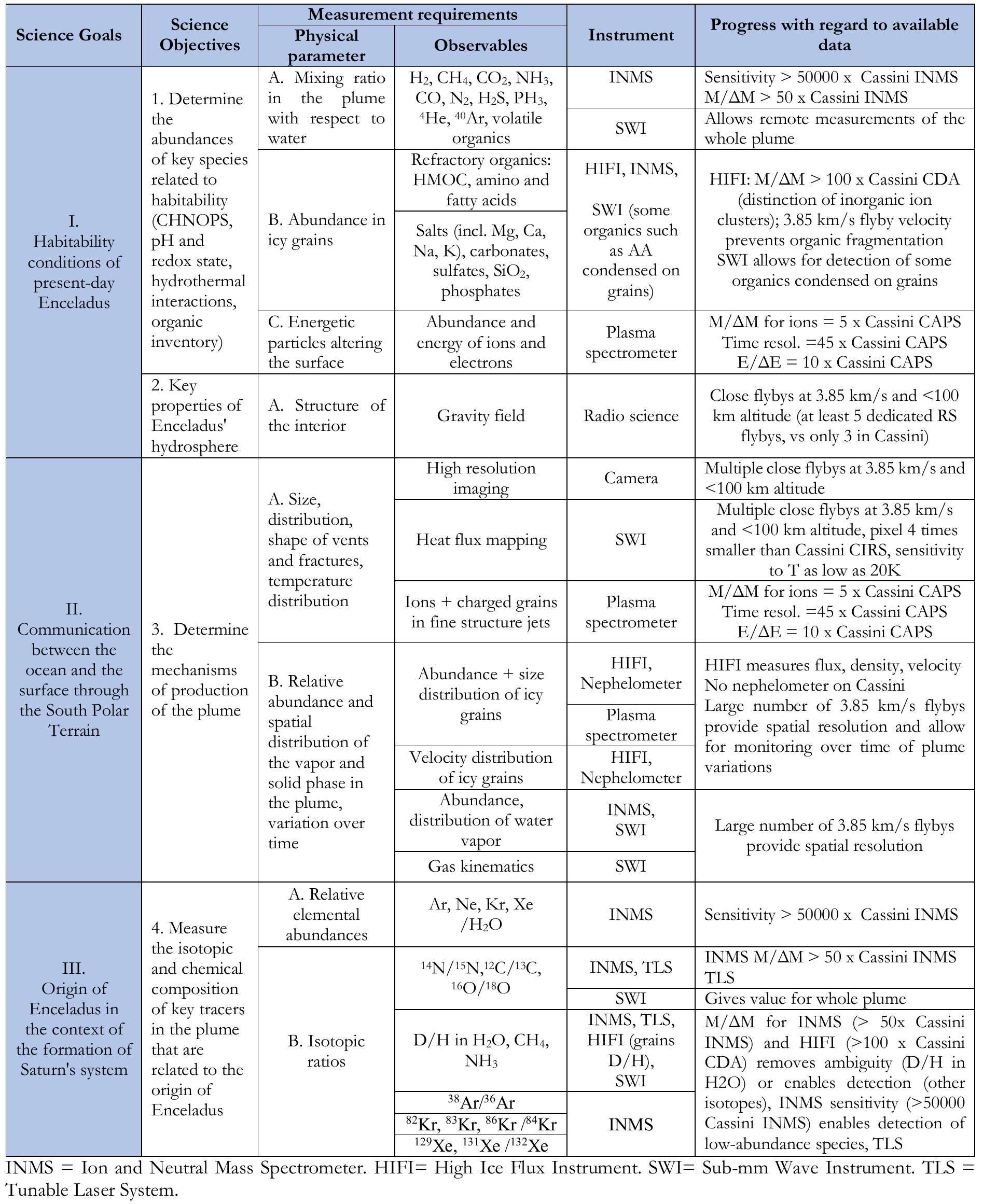} \\
\end{tabular}
\end{table*}

\section{Proposed Mission Configuration}
\label{sec:sec3}

In this Section, we first depict the proposed end-to-end {\it Moonraker} mission profile. We then provide a description of the spacecraft design, and finally discuss the needed technology requirements.

\subsection{End-To-End Mission Profile}

\subsubsection{Interplanetary Trajectory}
Our analysis has been performed assuming that the launcher for M-class missions would be an AR62 rocket, with a launch capability of $\sim$2650 kg for a departure at 3 km/s at the optimum declination (low-south latitude). A launch capability of $\sim$2185 kg for velocities in the 3.2 km/s range should be available up to a declination of $\sim$23$^\circ$ (N or S). The proposed baseline mission trajectory, considering a launch in March 2036 with a velocity of 3.23 km/s at a declination of 21.8$^\circ$ N, is then:

\begin{itemize}
\item EVEES trajectory: one swing-by of Venus (V), followed by a first Earth (E) swing-by to initiate a 3-years orbit, and a second Earth swing-by setting the spacecraft on a transfer trajectory to Saturn (S). No Deep Space Maneuver would be required;
\item Arrival at Saturn in May 2048, with a relative velocity of 5 km/s and a declination of 6.7$^\circ$.
\end{itemize}

Two back-up opportunities have been identified, with a launch in October 2036 and March 2038, with arrival dates at Saturn in mid-year 2048 and mid-year 2050, respectively. The proposed baseline mission would use the AR62 launcher. 

\subsubsection{Saturn Orbit Insertion and early tour phases}
Satellites of Saturn orbit close to the equatorial plane, with a large angle to its orbital plane (26.7$^\circ$). However, an arrival at Saturn less than 3 years after the Northern summer solstice (November 2045) results in declinations lower than 10$^\circ$ for the nominal and first back-up opportunities. A standard pump-down sequence 383 days long could then be implemented: 

\begin{itemize}
\item Titan flyby before Saturn Orbit Insertion (SOI) at an altitude of 1000 km, out of Titan’s atmosphere. This flyby reduces the magnitude of the Saturn Orbit Insertion maneuver, improving the mass budget. However, it may not be completely outside Titan's atmosphere, implying some uncertainty about the induced drag and trajectory perturbation, and a source of uncertainty for the SOI. Post flyby adjustment for the SOI trajectory might be needed;
\item SOI at pericenter (365000 km from the body center): $\sim$850 m/s for 5.25 km/s (explicit solution, the optimum requires adjusting the pericenter distance as a function of declination);
\item First orbit: 18:1 (287 days) with a large Pericenter Raise Maneuver (PRM) (250 m/s);
\item Second Titan flyby: 3.35 km/s (optimum for the science mission);
\item Pump down sequence at Titan: a series of orbits of decreasing periods -- 3, 2, 1 then 2/3 Titan periods (15.95 days) -- could be achieved with Titan flybys at altitudes higher than 1000 km (beyond the upper layers of the atmosphere of Titan). The Titan flyby following the 2/3 orbit sets the spacecraft to an orbit with a period close to 8 days, initiating the science phase with a first flyby of Enceladus.
\end{itemize}

\subsubsection{Science phase}
\begin{figure}[!ht]
\resizebox{\hsize}{!}{\includegraphics[angle=0,width=5cm]{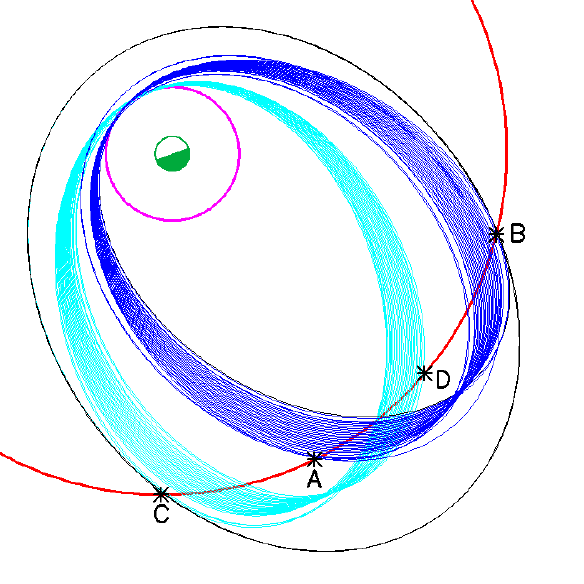}}
\caption{45 flybys of Enceladus distributed over two segments (1 segment for the nominal mission + 1 segment for an extended mission). Mission scheme: a first segment (dark blue, A: departure from Titan outbound; B: return to Titan inbound) provides 23 grazing flybys of Enceladus (magenta) over the South polar regions. For an extended mission, a 2:3 inbound-outbound orbit (black) could initiate a second segment (light blue) with 22 flybys shifted 30$^\circ$ clockwise; C: departure from Titan; D: return to Titan for disposal or transition to a new segment initiating a new extended mission.}
\label{fig:fig3}
\end{figure}

To perform multiple flybys over the South polar region of Enceladus, an orbital period in resonance with its 33 hours orbital period has to be selected. On this basis, we opted for an orbit with a 6:1 orbital period (8.22 days). This period is slightly larger than the 1:2 resonance with Titan (7.97 days). Every two orbits, Titan moves forward by 11.2$^\circ$, and more precisely 10.5$^\circ$ due to precession. After 30 orbits (157.5$^\circ$ forward motion), Titan begins to catch up with the alternate apocenter (180$^\circ$ of phasing away), but one could still implement more than 20 Enceladus flybys with minor Titan perturbation in between before Titan gets close to the apocenter again (see Figure \ref{fig:fig3}). A grazing encounter scheme has been selected, resulting in the slowest possible encounter velocity ($\sim$3.85 km/s), and minimizing the impact of precession on the encounters and orbit maintenance costs. Therefore, a mission segment would be constituted by a Titan to Enceladus transfer on an orbit with a period slightly larger than the 6:1 resonance after the first Enceladus flyby, leading to 21 to 24 Enceladus flybys. At the end of this segment, the spacecraft could be disposed of at Titan, or a new segment could be initiated using Titan flybys. Given the specific interest of the South high latitudes, all flybys have been set over the South polar region, with the first 2 flybys at an altitude of 200 km (to be compared to the $\sim$1500 km plume height), then the next flybys at 100 km or lower at the end of the nominal mission or during an extended mission. The latitude could however be varied to fit scientific requirements, e.g., to perform at least 5 dedicated radio-science flybys. The nominal {\it Moonraker} mission would have a duration of $\sim$13.5 years, with 189 days allocated for the science phase, assuming a 23--flyby segment.

Regarding end of mission, the Planetary Protection (PP) considers Enceladus as a Category III target, implying that this moon is not appropriate for crashing the spacecraft. {Only some specific landing conditions in the case of a descent module would be compliant with the Planetary Protection Policy (see \cite{Ne22} for details).} The proposed scenario would be spacecraft destruction during entry in Titan’s upper atmosphere. Other options to be considered would be crash on an intermediate-sized moon that has lower PP categorization, or delivery to a moon-free orbit around Saturn (strategy similar to Juno's end of mission at Jupiter). {Destruction in Saturn's atmosphere would be too expensive in terms of energetic cost.}

\subsubsection{$\Delta V$ budget for evaluating mass margins}

The deterministic and stochastic mission $\Delta V$ are found to be 1410 m/s and 118 m/s, respectively. The corresponding breakdowns are given in Table 2.

\begin{deluxetable*}{lll}
\tablecaption{Breakdowns of the deterministic and stochastic mission $\Delta V$.}
\tablecolumns{2}
\tablewidth{0pt}
\tablehead{Deterministic $\Delta V$ & }
\startdata 
Launch window 							& 50 m/s 													\\
Cruise 									& 200 m/s (440 m/s and 100 m/s for the two opportunities)		\\
SOI and pump down 						& 1100 m/s (to be improved after optimization)						\\
Nominal mission 							& 60 m/s (including disposal by collision with Titan after one segment)	\\
\hline
Stochastic $\Delta V$ 						& 														\\
\hline
Flyby clean-up during the interplanetary phase 		& 40 m/s 													\\
SOI + PRM 								&  30 m/s													\\
Titan flybys 								&  48 m/s  (8 m/s per flyby for JUICE)							\\
\enddata 
\label{tab:tab2}
\end{deluxetable*}

Clean-up costs would be negligible for Enceladus flybys given its very low gravity potential. With a velocity change of up to 10 m/s, a very large guidance error of 10 km only results in a difference of $\sim$0.3 m/s for the orbital velocity after the flyby. Flybys of Titan at an altitude of 10000 km or more would be implemented when initiating a segment. The velocity change is larger (400 m/s) than for Enceladus, but a 10 km guidance error also leads to a difference of $\sim$0.4 m/s. Therefore, only the 6 close flybys of Titan would be considered to require significant clean-up costs.

The total $\Delta V$ is then 1528 m/s (1604 m/s with a 5\% margin). {A two-segment} mission, for a total of 45 flybys of Enceladus over a 463 days mission, could be obtained for a deterministic $\Delta V$ cost of 150 m/s instead of 60 m/s, if remaining propellant allows. The following section considers only a one-segment mission (23 flybys).

\subsection{Spacecraft Design}

The structure of the platform would be mission-customized, with a strong heritage from JUICE. The design of the primary structure would be first driven by the central cylinder housing the propellant tanks and interfacing with the launcher, and the need to support the large solar panels as well as the JUICE-like 2.5 m diameter High Gain Antenna (HGA). This results in large support surfaces available to accommodate the payload instruments, the platform sensor and the other aerials, i.e., Medium Gain and Low Gain Antennae (MGA and LGA), thrusters’ pods and main engine. The propulsion would be similar to that of JUICE, but with downsized propellant tanks and a single pressurant tank, in line with mission $\Delta V$ and orbit control needs. The building blocks of our platform would be mostly inherited from JUICE. These elements have been considered in support of the mass and power budget consolidation. The main differences with respect to JUICE to meet the programmatic constraints of a M-class mission are:

\begin{itemize}
\item No dedicated NAVCAM as this functionality would be shared with the payload camera. As for JUICE, the processing algorithms of the NAVCAM images would run on the platform Command and Data Management Unit (CDMU);
\item {Two Star Tracker heads} instead of three on JUICE;
\item European Inertial Measurement Unit (IMU) Astrix 1000 family class would be operational and implemented in place of the US IMU of JUICE;
\item The communication system features only the X-band chain of JUICE. The steerable JUICE Medium Gain Antenna would be replaced by a smaller antenna supporting minimum communications at Saturn and a few kbps during the hot inner cruise of the early transfer phase between Earth and Saturn;
\item Due to the lower amount of solar and battery power to be managed, the digital conditioning system of JUICE Power Conditioning \& Distribution Units (PCDU) would be replaced by a standard analog Maximum Power Point Tracking (MPPT) system. A single battery module from JUICE instead of five meets the spacecraft (S/C) needs;
\item A couple of Input/Output (I/O) boards would be removed from the data management Remote Interface Unit, in line with the reduced number of interfaces to manage. The embedded science mass memory of the CDMU would be also of reduced capability, to match the downlink capability of the communication system operated at Saturn. 
\end{itemize}

Using the 3G28 or 3G30 solar cells of JUICE and considering the loss of efficiency (about 5\% due to majority carrier effect) in the colder operating environment, a 143~m$^2$ Mars Sample Return (MSR) array {(total area)} at 10 AU from the Sun delivers about 380 W (to be compared with 900W delivered by the JUICE 84 m$^2$ solar array at 5 AU), enabling an M-class spacecraft.  

Preliminary mass and power budgets are presented in Tables \ref{tab:tab3} and \ref{tab:tab4}. A payload of 50 kg conssuming 50 W over 3 hours, i.e., 150 Wh during each Enceladus flyby, would be allocated to the payload. System margins of 25\% (mass) and 30\% (power) are achieved on top of estimates. The energy budget would be balanced with 4 hours of communications per day considering a JUICE based system featuring a 2.5 m diameter HGA and the JUICE X-band telemetry chain. The amount of data downloaded to Earth varies between 0.9 Gb and 1.3 Gb between two Enceladus flybys separated by 8 days as a function of Earth/Saturn distance, to be allocated between the different instruments.

\startlongtable
\begin{deluxetable*}{l|c|c}
\tablecaption{ Mass budget derived from the JUICE avionics and MSR solar generator for a 50 kg science payload to Enceladus using A62 with 25\% system margin on the platform dry mass. Italic text denotes direct heritage from JUICE.}
\tablecolumns{3}
\tablewidth{0pt}
\tablehead{
Mass &   kg &    Heritage
}
\startdata 
Launcher capability & 2185 & Ariane 62 ($V_{\infty}$ = 3.3 km/s) \\
S/C max. launch mass & 2125 & Launcher capability -- 60kg adaptor\\
Propellant: && \\
- Orbit manoeuvers  & 850 & 1604 m/s (incl 5\% margin), 320 s Isp  \\ 
- Attitude control & 25 &  Allocation with margin \\
S/C max. dry mass & 1250 & \\
Payload & 50 & Allocation \\
System margin	& 250 &	25\% of S/C dry mass\\
 \hline
 Platform dry mass & 950 & Best estimate \\
 Structure &	185  &	15\% of S/C dry mass (best estimate)\\
 Solar arrays &	340 &	Mars Sample Return \\
 Harness	& 50	& 5\% of S/C dry mass\\
 Propulsion	& 120	& JUICE architecture with one pressurant tank and two 800 L bipropellant tanks \\
 Thermal	& 40 & Allocation\\
 AOCS &	70 & {\it 4 $\times$ reaction wheels}, {\it 2 STR}, 2 $\times$ Astrix IMU, {\it 2 Sun sensors}\\
 {Communication} &	60 & {\it X band system including HGA and 2 LGAs}, one 2 axes MGA 200 bps at Saturn\\
 Power &	60 & {\it 2 $\times$ battery modules}, 1 $\times$ customized PCDU, 1 axis Solar Array Drive Assembly\\
 DMS	&   25   & {\it CDMU} with limited SSMM, {\it RIU} with reduced I/O capabilities\\
\enddata 
\smallskip
{AOCS = Altitude and Orbit Control System. DMS = Data Management System}
\label{tab:tab3}
\end{deluxetable*}

\startlongtable
\begin{deluxetable*}{p{2.5cm}|p{2cm}|p{2.5cm}|p{3cm}p{2cm}|p{0.5cm}|p{2.5cm}}
\tablecaption{Energy budget of the mission achieved with the MSR solar array and 500 Wh battery range. The power system supports 3 hours of science at each flyby around Closest Approach (C/A) and 4 hours of communication per day with Earth.}
\tablecolumns{7}
\tablewidth{0pt}
\tablehead{
 Power &   Quiet cruise &    Communication sessions &   \multicolumn{2}{c|}{Science} &  &   \\
 &	 	& 		& 0.5 hour to 1.5 hour before and after C/A &  C/A $\pm$ 30 min 	&  		&  		
}
\startdata 
     Duration (h)	& 20.0	& 4.0	& 2.0	& 1.0 	& 	& 							    \\
     Solar arrays	& 380	& 380	& 380	& 0	    & W &  							    \\
     Total w/ margin & 353	& 503	& 380	& 489	& W & 							    \\
     System margin	& 81    & 116	& 88   & 113	& W	& 30\% Specified system margin  \\
     Net total	    & 272	& 387	& 293	& 377	& W & 								\\
     Payload	    & 0	    & 0	    & 50 	& 50 	& W & 								\\
     Propulsion	    & 5	    & 5	    & 5	    & 5	    & W &								\\
     Thermal	    & 70    & 50 	& 50 	& 50 	& W & 								\\ 
     AOCS	        & 75	& 75	& 100	& 180	& W &								\\
     Communication	& 35 	& 160	& 35    & 35 	& W & 								\\
     DMS			& 45 	& 50 	& 60	& 60 	& W & 								\\ 	 
 	 Power	        & 42 	& 47 	& 43 	& 47  	& W	& 95\% (30 W + regulation efficiency) \\ 
 	 \hline 
 	 Delta \% Solar arrays	&27	&-123	&0	&-489	& W & 								\\
 	 \hline
 	 Delivered to battery	& 26 	& 0	    &0	    & 0	    & W	    & 95\% BCR efficiency		\\
 	 Needed from battery    & 0		& 515   & 0     & 515   & Wh    & 95\% BCR efficiency  		\\
 	 Recharge capacity	    & 515   & 0	    & 0	    & 0 	& Wh	& Between communication session/after flyby	\\
\enddata 
\label{tab:tab4}
\smallskip
{BCR = Battery Charge Regulator.}
\end{deluxetable*}

\subsection{Technology requirements}
No new technology development has been identified for the spacecraft, in line with the requirements for an M--class mission. {Tailoring to the Saturn environment} is to be performed for the solar array and the solar cells. The absence of eclipses during the multiple orbits prevents the occurrence of surface temperatures as low as experienced on JUICE. The characterization first aims at securing power and energy budgets.

\section{Payload configuration}
\label{sec:sec4}
The scientific requirements discussed in Sec. 1 are addressed with a suite of scientific instruments listed in Table \ref{tab:tab5}. This list includes a Mass Spectrometer, a Tunable Laser System, a High Ice Flux Instrument, a Nephelometer, a Plasma Spectrometer, a Submillimeter Wave Instrument, a Camera, and a Radioscience Experiment. The total mass of the scientific payload would be 39.7 kg (with an allocation of 50 kg in the mass budget). This payload has been scaled to meet the specifications of an AR62 launch as provided by ESA. In case a more powerful AR version would be available for launch, the payload could be revised accordingly. The present configuration also enables additional science to be performed at Titan during the multiple approaches of the {\it Moonraker} spacecraft. 

\startlongtable
\begin{deluxetable}{p{5cm}|p{1.5cm}}
\tablecaption{Suite of scientific instruments.}
\tablecolumns{2}
\tablewidth{0pt}
\tablehead{
Instrument                              & Mass (kg) \\
}
\startdata 
{\it Moonraker} Ion \& Neutral Mass Spectrometer (M-INMS) & 6.2       \\
Tunable Laser System (TLS)              & 2         \\
High Ice Flux Instrument (HIFI)         & 4         \\
Nephelometer                            & 1         \\
Submillimeter Wave Instrument (SWI)     & 8         \\
Camera                                  & 12        \\
Radioscience Experiment                 & 2         \\
Plasma Spectrometer                     & 4.5       \\
Total Mass                              & 39.7      \\
\enddata 
\label{tab:tab5}
\end{deluxetable}

In the following, we provide a short description of each instrument, with their technology readiness and heritage.

\subsection{{\it Moonraker} Ion and Neutral Mass Spectrometer (M-INMS)} 
M-INMS measures the composition of the neutral gas and of the thermal ion population at the location of the spacecraft. In this configuration, the mass spectrometer would be provided by the University of Bern, Switzerland. This group has considerable space hardware experience, and in particular built the mass spectrometers RTOF and DFMS of the ROSINA experiment for Rosetta \citep{scherer2006novel,balsiger2007rosina}, the NGMS instrument for Luna-Resurs \citep{wurz2012neutral,hofer2015prototype}, and the NIM instrument of the PEP experiment on JUICE \citep{fohn2021description}. The proposed mass spectrometer is a time of flight (TOF) mass spectrometer. It thus measures a full mass spectrum at once with a mass resolution of $M$/$\Delta M$ up to 5000. {The nominal and extended mass ranges are 1--300 and 1--1000 amu.} The integration time for a mass spectrum {could} be adjusted between 0.1 and 300 s, to optimize the spatial resolution of the measurements as well as the sensitivity of the mass spectrometric measurements over a very wide altitude range along the flyby trajectory. The proposed mass spectrometer is comprised of an ion-optical system and an electronic box. The ion-optical system would be based on the RTOF instrument and the electronics on the NIM instrument on JUICE, which serves a very similar application (mass spectrometry during flybys of the Jovian moons). All subsystems have flight heritage, thus the combined system has a {Technology Readiness Level (TRL) of 6}.

\subsection{Tunable Laser System (TLS)} 
The instrument would be provided through a consortium composed of the Universities of Reims (GSMA-CNRS) and Aix-Marseille (LAM-CNRS). The instrument would be based on near-infrared antimonide laser diode absorption spectroscopy to provide concentration measurements of selected molecular species. The laser beam is propagated through the molecular gas where it is partially absorbed, the laser wavelengths being tuned to match accordingly with a rovibronic transition of the targeted molecules. The gas concentration is retrieved from the measurement of the amount of absorbed laser energy, using an adequate molecular model \citep{Zeninari06}. Hence, the sensor yields {\it in situ} measurements of gaseous abundances inside the plume of Enceladus, with a typical relative accuracy within a few {percent}. It would be an heritage of the former TDLAS instrument \citep{Durry10} launched within the framework of the Russian Martian mission Phobos-Grunt to provide measurements of C$_2$H$_2$, H$_2$O, CO$_2$ and their isotopologues \citep{Li09,Durry08,Barbu06a,Barbu06b} from the {\it in situ} pyrolysis of a Phobos soil sample. Therefore, the TRL of the TLS is around 8 \citep{durry2010near}. One should also mention that an upgraded version of the TDLAS developed and {led} by IKI/Roskosmos was ready to be launched within the framework of the Exomars-2022 Martian mission \citep{Rodin20}.

\subsection{High Ice Flux Instrument (HIFI)} 
HIFI would be provided by the Laboratory for Atmospheric and Space Physics at Colorado University. It would be an impact-ionization time-of-flight mass spectrometer which is optimized for measuring Enceladus plume ice grains. Plume grains strike an iridium impact plate at flythrough velocity, yielding a cloud of neutral and ionized species; HIFI measures the small resulting fraction of ionized species. Complementary cation and anion spectra allow determination of the composition of the salt-rich and organic-rich grains. The spectrometer has a $M/\Delta M$ of $\sim$3000, and would be able to detect organic and salt analytes in the grains’ icy matrix at ppm concentrations. {The considered mass range extends up to more than 2000 amu.} The HIFI instrument comprises two subsystems: a Mass Analyzer and an {Electronics} Box. The Mass Analyzer (MA) uses an electric field to extract ions created by impacting grains. The MA comprises two identical mass spectrometers: MA1 (cations) and MA2 (anions). The Electronics Box reads out the analog signals from the Sensor Head and controls its other components. HIFI would be the latest generation in a line of successful dust analyzer instruments: Giotto PIA at comet Halley \citep{kissel1986giotto}, Stardust CIDA at comet Wild 2 \citep{kissel2004cometary}, CDA Cassini at Saturn \citep{srama2004cassini}, and Europa Clipper SUDA (to be launched in 2024). It represents a large performance improvement over the similar ENIJA instrument \citep{srama2015enceladus} considered in the the E2T \citep{mitri2018explorer} and ELF \citep{reh2016enceladus} proposals. All HIFI subsystems are at TRL 6 and above, having been demonstrated in relevant environments via ground test of high-fidelity prototypes or operation in space; remaining integration involves standard interfaces and engineering development.


\subsection{Nephelometer}
The instrument, named LONSCAPE (Light Optical Nephelometer Sizer and Counter for Aerosols for Planetary Environments; \cite{renard20a}), would be provided through a consortium composed of the Universities of Orleans (LPC2E-CNRS) and Aix-Marseille (LAM-CNRS). The instrument provides the scattering function at several angles of the particles that cross a laser beam inside an optical chamber. By doing so, LONSCAPE performs measurements of the concentrations and the typologies of the particles for 20 size classes in the 0.1--30 $\mu$m range. The counting is performed at the small scattering angles, where the scattered light is mainly dependent on the diffraction and thus {not} sensitive to the refractive index. The typology is retrieved from the scattering properties at the several angles, by comparison with laboratory measurements. The team has developed instrumentation for particles detection in particular under stratospheric balloons \citep{Renard20b} and in nanosatellites \citep{Verdier20}, with current TRL 6-7 for space applications. Also, some studies have been conducted for an application to the aerosols detection in the upper atmosphere of Venus \citep{Baines21}.

\subsection{Submillimetre Wave Instrument (SWI)} 
The Submillimetre Wave Instrument (SWI) would be provided by a consortium led by the Max Planck Institute for Solar System Research (MPS)\footnote{https://www.mps.mpg.de/planetary-science/juice-swi}, and corresponds to a passively cooled tunable heterodyne spectrometer covering the frequency range of 1065 to 1275 GHz. The local oscillator chain consists of a 25 GHz band frequency synthesizer. Its output signal is tripled to 75 GHz where it is amplified with an E-band amplifier. Three frequency doublers produce a signal of a few mW at 600 GHz, {feed} to a subharmonically pumped mixer. The spectrometer backend consists of a Chirp Transform Spectrometer with 1 GHz bandwidth and 100 kHz spectral resolution. The resolving power of the instrument is above 1 $\times 10^7$. The receiver is coupled to a telescope with a 29 cm primary mirror. The spatial resolution of the telescope is about 1 mrad. SWI would be based on the JUICE-SWI instrument, mounted into the JUICE satellite in August 2021, however due to mass constraints and different scientific objectives of {\it Moonraker} compared to JUICE, a number of components  would be  descoped: the 600 GHz receiver, the along and cross track actuators and the autocorrelator spectrometers. All components of the {\it Moonraker} SWI are TRL 8.  
 
 
\subsection{Camera}
The camera for {\it Moonraker} would be provided by a partnership between the University of Aix-Marseille (LAM-CNRS) and the Institute for Planetary Research of DLR. DLR, LAM and their partners have many years of experience in design of optical imaging instruments and their key-components for planetary science missions (Mars-Express (PI), Rosetta-Lander (PI), DAWN, Hayabusa-II, ExoMars (PANCAM-HRC) and recently JUICE (Co-PI)). The camera would be a straightforward telescope combined with a scientific VIS/NIR CMOS- image sensor (based on JUICE-JANUS) and their associated electronics \citep{De14}. {The typical angular resolution that could be achieved is 15 $\mu$rad/pixel. Because the SPT is expected to be in the dark of polar winter during the science phase, the illumination due to Saturn's glow will have to be quantified to provide a better assessment of the camera design. To do so, a wide dynamic range is required.} The key components of the camera are TRL 8. In our spacecraft design, this camera would be also used as a navigation camera.

\subsection{Radioscience Experiment}
Doppler tracking of the spacecraft during flybys (preferably via the HGA) would provide the determination of the gravity field and the orbits of the moons \citep{Ie14,Du19}. In the {\it Moonraker} mission concept, gravity is the only available tool to constrain the interior structure of the Saturnian satellites. The precise knowledge of Enceladus’ orbit is crucial to characterize the dissipative processes in the Saturnian system. The Doppler data would be produced at the ground antenna via a 2-way coherent link. The measurements use the onboard Radio Communication System (RCS), without the need for dedicated equipment. If the RCS includes the Integrated Deep Space Transponder (IDST) developed by ASI and ESA, to be flown on NASA’s VERITAS mission to Venus, Ka-band tracking becomes possible, providing a significant enhancement of the data quality and the determination of the interior structure. Indeed, Ka band radio links (32.5--34 GHz) are nearly immune to plasma noise, the main limitation to Doppler measurements in X band tracking systems (7.2--8.4 GHz). The IDST could be a contribution of the Italian Space Agency to Moonraker, or be provided by ESA as the onboard transponder, a system element. The IDST, which would be based on the digital technologies developed for BepiColombo MORE investigation, enables also range measurements accurate to 1--4 cm \citep{Ca20}, thus providing very precise data on the orbits of the moons. 

\subsection{Plasma spectrometer} 
The instrument would be provided by a consortium under the responsibility of the University of Toulouse (IRAP-CNRS). The formed international consortium has a solid institutional experience and outstanding mission heritage, including BepiColombo/MEA and MSA, MAVEN/SWIA, JUICE/JDC, and Cassini/CAPS \citep{Yo04,de16,sa10,sa21,Wi19} to thoroughly address the scientific objectives of the Plasma Spectrometer. The instruments consist of an electron and negative ion spectrometer (1--30 keV/q) together with an ion mass spectrometer (1--40 keV/q, $M/\Delta M$=40 for $<$15 keV/q) with shared LVPS and DPU. The current TRL for the Plasma Spectrometer is 5 at minimum for all its elements and would be planned to reach TRL 6 by the end of Phase A. 

\section{Management Structure}
\label{sec:sec5}

The {\it Moonraker} mission concept is proposed as an ESA-led mission, with a contribution to the science payload by NASA. Participating in the elaboration of the {\it Moonraker} proposal is one industrial company, Airbus Defense and Space. The international Consortium for the {\it Moonraker} mission concept involves the platform as well as the science instruments and science investigations. After selection by ESA, the European industrial partner would be responsible for developing, within the international Consortium, the platform. The {\it Moonraker} instrument payload is provided by instrument PI teams from ESA’s Members states and NASA scientific communities. Payload funding for ESA’s members states is provided by National funding agencies, while the U.S. payload contribution would be funded by NASA. The lead-funding agency for each PI-team is either the PI National Funding Agency for a European PI-led team or NASA for a U.S.-led PI team.

\section{Conclusion}
\label{sec:sec6}

The {\it Moonraker} mission concept has been submitted to the ESA Call for a medium-size mission opportunity released in December 2021. It consists of an ESA-provided platform, with strong heritage from JUICE and Mars Sample Return, and carrying a suite of instruments dedicated to plume and surface analysis. The nominal {\it Moonraker} mission concept includes a 23--flyby segment and has a duration of $\sim$13.5 years, with 189 days allocated for the science phase. It can be expanded with additional segments, if needed, to satisfy the science objectives. The ESA review indicated that the needed budget is larger than the maximum one at disposal for medium size missions (550 millions euros), and that the mission profile is rather tailored to match that of a large-size missions in terms of budget ($\sim$one billion euros) and mission design (need of a rocket more powerful than AR62, 12--year cruise duration, extended international collaboration, and significant operation duration). 

The submission of an extended {\it Moonraker} proposal is currently envisaged to the next ESA Call for a large-size mission, including science both at Enceladus and Titan, which would fit the ``Moons of the giant planets'' priority defined for L-class missions in the Voyage 2050 program. The {\it Moonraker} mission concept corresponds to one of the top priorities of the future New Frontiers 6 and 7 calls proposed by the {2023--2032 US National Academies' Planetary Science and Astrobiology Decadal Survey}. Enceladus is also considered as the second highest priority new Flagship mission for the decade 2023-2032 recommended by this panel, with the highest priority attributed to the Uranus Orbiter and Probe (UOP). 

In case Ariane 64 (AR64) could be envisaged, additional mission capability would be considered (e.g., extension of the payload such as the addition of a magnetometer, more propellant to increase the orbital phase duration).  The additional available launch mass could be considered to piggyback an additional contribution (to explore the Saturn system) from another space agency. 

If selected, such a spacecraft would also provide important follow-up science at Titan after the Dragonfly mission. Flybys of Titan are already a part of the initial mission profile, although at a high altitude. The additional fuel available could allow for closer flybys, enabling mass spectrometry measurements in Titan's atmosphere. {Such a mission concept would have to overcome any potential contamination between the successive flybys of Titan and Enceladus. To do so, M-INMS could contain a bake-out heater designed to heat the ion source up to 150--300$^\circ$C for at least 24 h. This ion source bake-out heater would clean the ion source from any contaminant deposited there and originating from the spacecraft. It would also remove any chemical heritage from prior measurements, such as the Titan atmosphere. Such an ion source bake-out system is standard for neutral gas mass spectrometers, as illustrated by the ROSINA/Rosetta experiment, but also by the Neutral Gas Mass Spectrometer (NGMS) instrument for Lunar Resurs, and the Neutral Ion Mass Spectrometer (NIM) on the Particle Environment Package (PEP) of JUICE. If the ion source bake-out heater would still be considered insufficient, then a hermetically sealed instrument  could be flown, with the seal only broken at Enceladus. In addition, the risk of contamination of the HIFI instrument would be almost zero given the fact that it is more or less in a vacuum tight housing, with a tiny aperture and a target fully isolated from the ambient atmosphere. The HIFI team is also currently investigating the possibility to keep the target warm during flybys to prevent ice grains from sticking and slowly evaporating.}

A larger science payload should be considered, with the inclusion of a radar and a magnetometer. {The addition of a magnetometer would allow us to use the reported tidal variations in plume activity to explore Enceladus' inductive response to the resulting time-dependence of the background magnetic field arising from the moon-magnetosphere interaction. This will place a constraint on global ocean thickness and salinity.} This instrument would also allow for additional science related to Saturn's magnetosphere and Titan's ionosphere. An ice-penetrating radar would also allow for further understanding of the communication of Enceladus' ocean with its surface, as well as help establishing bathymetry maps of Titan's lakes and seas. An additional module to be dropped at Titan, such as a small entry probe or a minisatellite, could be considered as well.

\begin{acknowledgments}
OM and AB acknowledge support from CNES. JIL was supported by the JPL Distinguished Visiting Scientist Program. ZM acknowledges funding from FEDER--Fundo Europeu de Desenvolvimento Regional funds through the COMPETE 2020--Operational Programme for Competitiveness and Internationalisation (POCI), and by Portuguese funds through FCT--Fundação para a Ciência e Tecnologia in the framework of the project POCI-01-0145-FEDER-029932 (PTDC/FIS-AST/29932/2017).  Centro de Química Estrutural acknowledges the financial support of FCT- Fundação para a Ciência e Tecnologia (UIDB/00100/2020 and UIDP/00100/2020), and Institute of Molecular Sciences acknowledges the financial support of FCT—Fundação para a Ciência e Tecnologia (LA/P/0056/2020). Some of this work was conducted at the Jet Propulsion Laboratory, California Institute of Technology, under a contract with the National Aeronautics and Space Administration (80NM0018D0004). Reference herein to any specific commercial product, process, or service by trade name, trademark, manufacturer, or otherwise, does not constitute or imply its endorsement by the United States Government or the Jet Propulsion Laboratory, California Institute of Technology. In Memory of Prof. Anny-Chantal Levasseur-Regourd.

\end{acknowledgments}
\bibliography{biblio}{}
\bibliographystyle{aasjournal}

\end{document}